\documentclass[preprint]{aastex}

\def\be#1{\begin{equation}\label{eq:#1}}
\def\ee{\end{equation}}
\def\EC#1{(\ref{eq:#1})}
\def\bea#1{\begin{eqnarray}\label{eq:#1}}
\def\ee{\end{equation}}
\def\eea{\end{eqnarray}}
\def\iatrous{$\it{algorithme~\grave{a}~trous}~$}
\def\iatrousp{$\it{algorithme~\grave{a}~trous}$}
\newcommand{\bfatrous}{\mbox{\boldmath $\bf{algorithme~\grave{a}~trous}~$}}
\newcommand{\bfAtrous}{\mbox{\boldmath $\bf{Algorithme~\grave{A}~Trous}~$}}

\begin{document}

\title{Multiresolution Analysis of Substructure 
in Dark Matter Halos}

\author{Michael D. Seymour\altaffilmark{1} and 
Lawrence M. Widrow\altaffilmark{2}}
\affil{Department of Physics, Queen's University, 
Kingston, Ontario, Canada K7L 3N6}

\altaffiltext{1}{seymour@astro.queensu.ca}

\altaffiltext{2}{widrow@astro.queensu.ca}

\begin{abstract}

Multiresolution analysis is applied to the problem of halo
identification in cosmological N-body simulations.  The procedure
makes use of a discrete wavelet transform known as the \iatrous
and segmentation analysis.  It has the ability to find
subhalos in the dense regions of a parent halo and can discern the
multiple levels of substructure expected in the hierarchical
clustering scenario.  As an illustration, a 500,000 particle dark
matter halo is analyzed and over 600 subhalos are found.  Statistical
properties of the subhalo population are discussed.

\end{abstract}


\section{Introduction}

The hierarchical clustering hypothesis provides an attractive paradigm
for the formation of structure in a universe dominated by cold dark
matter.  Small-scale objects form first and merge to yield systems of
increasing size.  This highly non-linear process has been studied
extensively using N-body simulations with particular attention paid to
the survival of subhalos once a merger event has occurred.  Early
results suggested that substructures (i.e., subhalos within halos) are
erased efficiently (White 1976; Frenk et al.\,1988).  This so-called
overmerging problem plagued investigations of galaxy cluster formation
since it lead to the conclusion that the constituent galaxies do not
survive.  However, recent high resolution simulations (Ghigna et
al.\,1998, Klypin et al.\,1999) together with analytic work (Moore,
Katz, \& Lake 1996) demonstrated that the overmerging problem was due
entirely to the poor mass and spatial resolution of early simulations.
Indeed, Moore et al.\,(1999) found that on galactic scales, simulated
halos may have too much substructure.  Their high resolution
simulation of a $10^{12}M_\odot$ (i.e., Milky-Way sized) halo revealed
over 500 $M\ga 10^8\,M_\odot$ subhalos, a factor of 50 greater than
the number of visible satellites in the Milky Way.

An essential element in the analysis of cosmological N-body
simulations is the identification of physical structures, namely halos
and subhalos.  There are now a number of algorithms available to do
this such as friends-of-friends (FOF; Davis et al.\,1985), DENMAX
(Bertschinger \& Gelb 1991; Gelb \& Bertschinger 1994), and SKID
(Governato et al.\,1997).  The FOF algorithm identifies structures by
linking all particle pairs separated by less than a user-supplied
distance known as the linking parameter.  For a particular choice of
linking parameter, the algorithm produces a unique list of structures.
DENMAX, and the closely related algorthm SKID, are based on contour
surfaces of the three-dimensional density field.  Each local maximum
of the density field is assumed to correspond to the center of a halo
or subhalo: All particles interior to the last closed contour
surrounding a given maximum are assigned to the corresponding halo.

Two challenges, brought to the fore by the dramatic improvements in
the mass and force resolution of present-day simulations, now confront
these methods: (1) identification of subhalos within the dense regions
of a parent halo; and (2) analysis of multiple levels of substructure.
A given simulation particle may be a member of a small clump that is,
in turn, gravitationally bound to a larger (sub)halo.  Likewise, the
subhalo may be gravitationally bound to a galactic or cluster-sized
halo.  Such a particle is most accurately described as being a member
of three distinct structures.  Thus, any scheme which assigns a given
particle to at most one structure cannot hope to capture the
hierarchical nature of dark matter halos.

Early incarnations of the FOF algorithm relied on a single linking
length.  If the linking length is set to be too large, subhalos in the
inner parts of large halos are missed.  With a small linking length,
on the other hand, the algorithm picks out substructure but loses
information on large scales.  These problems can be avoided if one
uses a ``hierarchical'' version of FOF (Klypin et al.\,1999) wherein
the algorithm is run several times with different linking lengths.
DENMAX and SKID are best suited to finding substructure since they
locate all maxima in the density field.  The hierarchical nature of
clustering can be studied by applying smoothing filters to the density
field and rerunning the algorithm.

In this paper, we describe a multiresolution analysis (MRA) that
handles, in a natural way, the multiple levels structure found in
cosmological N-body simulations.  MRA refers to a general class of
tools that provide a simple hierarchical representation of a signal,
the signal, in our case, being the density field in a simulation.  At
each resolution, the analysis picks out the {\it details} of the
signal at a characteristic scale.  Thus, MRA can be thought of as a
``mathematical microscope'': Coarse resolution (low magnification)
probes large-scale structures in the signal while fine resolution
(high magnification) probes small-scale structures.

We employ a specific MRA that is based on the wavelet transform known
as the \iatrousp.  A wavelet transform allows one to analyze a signal
simultaneously in scale and position.  The transform accomplishes this
task by convolving the signal with a special type of window function
known as a wavelet.  Wavelets must have compact support and integrate
to zero.  The wavelet transform therefore probes local properties of
the density field and is insensitive to a mean background.  The
wavelets used in a particular implementation of the transform are
chosen to be translations and dilations of a single prototype known as
the mother wavelet.  A low resolution analysis of a signal is achieved
by using large-scale versions of the mother wavelet: high resolution
is achieved by using small-scale versions.  Thus, wavelet analysis
fits naturally within the framework of MRA (Mallat 1989a, 1989b).

The wavelet transform can provide a complete description of a density
field found in a cosmological N-body simulation.  Segmentation
analysis produces, from the results of the wavelet transform, a
catalog of structures and substructures.  Dynamical information can
then be used to cull, from this catalog, unbound associations of
particles.

The algorithm described in this paper was first used in the context of
cosmological N-body simulations by Lega et al.\,(1995) and Lega et
al.\,(1996) who were interested in the morphology of large scale
structure in various cosmological models (e.g., hot vs.\,cold dark
matter).  It has also been used by Gambera et al.\,(1997) to analyze
observational data for the Coma cluster.  We adapt this algorithm to
the specific task of sorting out multiple levels of substructure that
arise within individual dark matter halos.  Our implementation of the
algorithm is outlined in Section 2.  Results from the analysis of a
cluster-sized dark matter halo are presented in Section 3.  In Section
4 we describe our somewhat unorthodox procedure for identifying bound
systems of particles.  Statistics of the subhalos are discussed in
Section 5 where we also present an example of a region of the
simulation that exhibits multiple levels of substructure.

\section{Multiresolution Analysis of an N-body Simulation}

In this section, we describe the two steps which comprise the heart of
the algorithm: (1) calculation of the wavelet transform of a density
field; and (2) application of segmentation analysis to identify 
individual physical structures.

\subsection{From the Continuous Wavelet Transform to the 
\bfatrous}

The wavelet transform of a given signal represents the detail in the
signal as a function of position and scale.  Operationally, the
wavelet transform is computed by convolving the signal with a window
function known as a wavelet.  For a particular wavelet transform, the
set of analyzing wavelets is constructed by applying translations and
dilations to the mother wavelet.  Thus, the mother wavelet
characterizes the transform.

In a continuous wavelet transform (CWT), the set of analyzing wavelets
is chosen to include all possible translations and dilations of the
mother wavelet.  If the signal depends on one variable (say position)
the CWT will be a continuous function of two variables, position and
scale.  It follows that the information contained in a CWT is highly
redundant.  By contrast, the discrete wavelet transform (DWT) utilizes
a countable set of analyzing wavelets constructed by taking discrete
translations and dilations of the mother wavelet.  Provided certain
conditions are met, the redundancy in the wavelet transform can be
eliminated.  In other words, the set of analyzing wavelets in a DWT
can be chosen to form a complete, orthonormal basis for a large class
of continuous functions.

For large data sets, the number of computations required in the
standard DWT becomes exceedingly large.  There are now a number of
efficient algorithms which make it possible to perform DWTs on large
data sets.  These algorithms are related to the standard DWT much in
the way the fast Fourier transform is related to the standard Fourier
transform.  The \iatrous (Holschneider et al.\,1989, Dutilleux 1989)
is an example of such an algorithm.  Our description of this algorithm
is preceded by a general discussion of continuous and discrete wavelet
transforms.

The CWT of a one-dimensional function $f(x)$ is given by

\be{wavelet} 
w\left (\lambda,\,a\right ) =
\int_{-\infty}^{\infty}f(x)\, \psi^{*}\left (\lambda,\,a;\,x\right )dx 
\ee

\noindent where the parameters $\lambda$ and $a$ characterize the
scale and position of the analyzing wavelet $\psi\left
(\lambda,\,a;\,x\right )$ and $\psi^*$ is the complex conjugate of
$\psi$.  $\psi\left (\lambda,\,a;\,x\right )$ is constructed by
applying a rescaling (by $\lambda$) and a translation (by $a$) to the
user-supplied mother wavelet $\psi_0$:

\be{wavelets_def} 
\psi\left (\lambda,\,a;\,x\right )~=~
\frac{1}{\sqrt{\lambda}} \psi_0\left (\frac{x-a}{\lambda}\right )~.
\ee 

\noindent
The original function $f$ can be reconstructed via the inverse
transform

\be{inverse}
f(x) = \int\int w\left (\lambda,\,a\right )
\psi\left (\lambda,\,a;\,x\right )d\lambda\,da
\ee

\noindent
provided the mother wavelet satisfies the admissibility condition

\be{admiss}
\int\,\left |{\hat\psi}_0(k)\right |^2
\,\frac{dk}{|k|}~<~\infty
\ee

\noindent
where ${\hat\psi}_0$ is the Fourier transform of $\psi_0$.  From
this condition it follows that ${\hat\psi}(0)\propto\int
dx\,\psi(x)=0$ and therefore the wavelet must be oscillatory, i.e.,
wavelike in nature.

A CWT resembles a windowed Fourier transform (WFT) in the sense that
both operations utilize a set of window functions or filters to
analyze a signal simultaneously in scale and position (see, for
example, Kaiser 1994; Burrows, Gopinath, \& Guo 1998).  However, the
width of the window function in the CWT scales with $\lambda$ whereas
the width of the window function in a WFT is fixed.  Thus, the CWT is
{\it scale-independent}: Rescaling lengths in the data set leaves the
CWT unchanged modulo a shift along the scale axis.

The CWT defined in Eq.\,\EC{wavelet} maps a function of one variable
into a function of two variables.  Thus, the resultant transform, $w
\left (\lambda,\,a\right )$, contains redundant information (i.e.,
$\psi\left (\lambda,\,a;\,x\right )$ forms an overcomplete basis).
Consider instead the DWT and associated reconstruction formula:

\be{DWT2} 
w_{n,i} = \int_{-\infty}^\infty \psi_0^*\left (
\frac{x-a_i}{\lambda_n}\right )\,f(x)
\ee

\be{DWT}
f(x) = \sum_{n,i} 
\psi_0\left (\frac{x-a_i}{\lambda_n}\right )
\,w_{n,i}
\ee

\noindent where $i$ and $n$ are integers, $a_i = i\Delta a$, and
$\lambda_n = \sigma^n Delta a$.  $\sigma$ and $\Delta a$ are constants
representing respectively the ratio between different levels $n$ of
the DWT and the sampling interval.  The ${w_{n,i}}'s$ are referred to
as the wavelet amplitude coefficients (WACs).  In what follows, we
take $n=1$ to correspond to the smallest scale (highest resolution)
level in the transform.  Increasing $n$ corresponds to dilating the
wavelet and thus degrading the resolution.  In practice, the signal is
sampled at regular intervals --- for the case at hand, the sampling
interval corresponds to the grid spacing of the discretized density
field --- and it is natural to choose $\Delta a$ to be equal to the
grid spacing.  The integral in Eq.\,\EC{DWT2} is then replaced by a
sum over $j$ where $x\Rightarrow x_j=j\Delta a$.

The goal in wavelet analysis is to find a set of analyzing wavelets
such that any square-integrable function can be expressed by
Eq.\,\EC{DWT}.  Many examples of suitable wavelet bases have been
discovered.  However, for large data sets, the computation
requirements become prohibitively large.  Consider a signal that
consists of $N$ data points.  If the wavelet has compact support so
that it takes $m$ computations to perform a single convolution at
level $n=1$, it will take $m\times \sigma^{(n-1)}$ computations to
perform a convolution at level $n$ and therefore $N\times m\times
\sigma^{(n-1)}$ computations in total for level $n$ of the transform.
In principle, we may consider wavelets whose scale is comparable to
the size of the data set, i.e., $m\times \sigma^{{n_{\rm max}}-1}\sim
N$ where $n_{\rm max}$ refers to the highest level of the transform.
The number of computations for this level will be $O(N^2)$.  Since
$n_{\rm max}\simeq \log_\sigma\left (N/m\right )$ the complexity of
the transform will be $N^2\log_\sigma\left (N/m\right )$.

The \iatrous is an efficient scheme for performing discrete wavelet
transforms.  The algorithm assumes a scaling parameter $\sigma=2$.
Wavelet coefficients are obtained recursively in that the WACs at
level $n$ are calculated by taking an appropriate sum over WACs at
level $n-1$.  Thus, the number of computations necessary for the
highest level of the transform is at most $O(N)$ (or more accurately,
$O(N\ln{N})$).  Since the algorithm is an example of the application
of wavelets to MRA (Mallat 1989a, 1989b) we first interject a few
general remarks about MRA.

Consider a function $f(x)$ sampled at a fixed regular interval $\Delta
a$.  This highest resolution sample of $f$, which we refer to as
$f_0$, is constructed by convolving $f$ with a window function $\phi$:

\be{smooth1}
f_{0,i} = \sum_j \phi\left (x_i-x_j\right )f\left (x_j\right )~.
\ee

\noindent (The second subscript on $f_0$ refers to position, i.e.,
$x_i = i\Delta a$, etc.).  Smoother representations of $f$ can be
constructed with filters $\phi_n$ that are dilated versions of $\phi$:

\be{smooth2}
f_{n,i} = \sum_j \phi_n\left (x_i-x_j\right )f\left (x_j\right )
\ee

\noindent where

\be{recursion2}
\phi_n \left(x\right ) = \frac{1}{2^n}\phi\left (\frac{x}{2^n}\right )~.
\ee

\noindent For this reason, $\phi$ is referred to as the scaling
function.

The ``details'' $w_{n,i}$ of the signal correspond to the part of
$f$ that is removed as one degrades the resolution from one level to
the next.
Schematically, we have

\be{sumforf}
f_0~=~w_1+w_2+\ldots w_{N-1}+ f_N 
\ee 

\noindent where $f_N$ is the lowest resolution version of $f$ and 

\bea{details} 
w_{n,i} &\equiv & f_{n-1,i}-f_{n,i} \nonumber \\
& = &  \sum_j \left (\phi_{n-1}\left (x_i-x_j\right ) -
\phi_{n}\left (x_i-x_j\right )\right ) f_{x_j}~. 
\eea 

\noindent Comparing Eqs.\,\EC{DWT} and \EC{details} we see that the
mother wavelet can be calculated by a simple subtraction

\be{waveletdef} 
\psi(x) = \phi(x)-\frac{1}{2}\phi\left (\frac{x}{2}\right )~. 
\ee 

\noindent For MRA, we require that the scaling function satisfies the
recursion relation

\be{recursion1} 
\phi\left (\frac{x}{2}\right ) = 
\sum_j h_j \phi\left (x-x_j\right ) 
\ee 

\noindent where $h_j$ are constants (Mallat 1989a, 1989b; see, also
Kaiser 1994).  It is this equation that enables us to calculate the
WACs recursively.

The \iatrous (Holschneider et al.\,1989, Dutilleux 1989) is an MRA
that has a number of attractive features.  The algorithm is
computationally efficient and easy to program.  For example,
reconstruction of the original function involves a simple sum over
scale at each position.  In three dimensions, the algorithm is
approximately isotropic.  Finally, the wavelet coefficients at each
level are calculated for all points of the highest resolution
(interval $\Delta a$) grid and therefore the shapes and sizes of
structures are well-determined.  By contrast, the original MRA
developed by Mallat (1989a, 1989b) is computationally intensive, the
reconstruction formula is complicated, and the transform in 2 or more
dimensions is not isotropic.  Moreover, wavelet coefficients are
calculated on a decimated grid (in going from level $n$ to level $n+1$
the number of wavelet coefficients is reduced by a factor of 2 in one
dimension or 8 in three dimensions) and therefore the representation
of larger structures becomes rather crude.  A detailed comparison of
Mallat's MRA and the \iatrous can be found in Shensa (1992).

Following Lega et al.\,(1995), we choose the scaling function to be
the cubic B-spline:

\be{aux_function}
\phi(x)=\frac{1}{12}\left (
|x-2|^3-4|x-1|^3+6|x|^3-4|x+1|^3+|x+2|^3\right )
\ee

\noindent 
where distances are measured in units of $\Delta a$.  Note that
$\phi(x)=0$ for $|x|\ge 2$.  The recursion formula,
Eq.\,\EC{recursion1}, takes the form

\be{recursion3}
\phi\left (\frac{x}{2}\right ) =
\frac{1}{8} 
\sum_{j=-2}^2 C_{2-j}^4\phi\left (x-x_j\right )
\ee

\noindent where $C_n^m\equiv n!/m!(n-m)!$ are the usual binomial
coefficients.  Eq.\,\EC{recursion3} can be verified by a
straightforward but tedious calculation.  The mother wavelet,
calculated from Eq.\,\EC{waveletdef} and shown in Figure 1, is similar
to the popular Mexican Hat wavelet (the Laplacian of a Gaussian).
Explicit formulae used in the calculation of the wavelet coefficients
are given in Appendix A.

As discussed above, $\int dx\,\psi_0(x) = 0$.  In addition, the first
moment of $\psi_0$ also vanishes: $\int dx \,x\,\psi_0(x) = 0$.  It
follows that the leading contribution to the wavelet transform of
$f(x)$ is from terms quadratic in $x$: The wavelet transform is
evidently unaffected by a constant or linearly varying background.  It
is this property of the wavelet transform that makes it ideal for
identifying structures in cosmological simulations.

In three dimensions, the wavelet is constructed from the scaling
function $\Phi$, assumed to be separable in Cartesian coordinates:

\be{auxiliar3D} 
\Phi\left ({\bf r}\right )\equiv \phi(x)\phi(y)\phi(z)~.
\ee 

\noindent
Once again, explicit formulae for the WACs are given in Appendix
A.  The wavelets constructed in this manner are quasi-isotropic, i.e.,
the profile of the mother wavelet depends on the direction in space,
though it is the same along any three mutually orthogonal directions.
Moreover, the differences along non-orthogonal directions are
relatively minor as can be seen in Figure 1 where we compare the
profile of the wavelet along one of the three original axes with that
along one of the diagonals $x=\pm y= \pm z$.

\subsection{Application of the \bfAtrous to N-body Simulations}

The first step in applying the \iatrous to an N-body simulation is to
discretize the density field by approximating the density at each
point of a three-dimensional Cartesian grid.  We use the
nearest-grid-point scheme frequently employed in particle-mesh
simulations (Hockney \& Eastwood 1988).  Simply put, the density at a
particular grid point is given by the sum of the masses of all
particles within a cell centered on that point divided by the volume
of the cell.

For the analysis presented in this paper was use a grid $128^3$ in
size, the maximum one allowed given memory constraints of the
available computers. The number of grid points is roughly a factor of
four greater than the number of particles in the simulation and
therefore we do not expect that a finer mesh will lead to an
improvement in the results.  Indeed, the highest resolution wavelets
are able to pick out associations of very small numbers of particles.

The $n=1$ wavelet has a width of approximately five cells.  Thus, only
wavelets up to $n=5$ (80 cells across) fit into our grid.  The $n=5$
level essentially provides a map of the gross features of the halo and
therefore useful information on subhalos is contained in levels
$n=1-4$.

With the discretized representation of the density field in hand, the
WACs can be calculated according to the formulae in Appendix A.

\subsection{Segmentation Analysis}

The multiresolution analysis described above provides a complete
description of the (discretized) density field obtained in an N-body
simulation.  The next step in the analysis is to develop a
prescription for identifying physical structures.  For this step, we
take, as a working definition of structure, an association of
particles in space that is unlikely to have occurred by chance.  This
definition includes both gravitationally bound clumps and the debris
of systems that have been tidally disrupted by the parent halo.  In
the next section, we describe how one can distinguish between these
two possibilities.

In the wavelet description, structures are identified as connected
regions where the WACs rise above a predetermined threshold.  The
introduction of a user-supplied threshold is a feature common to all
clump-detection algorithms.  For example, in the FOF algorithm, the
threshold corresponds to the linking parameter.  In an MRA, a different
threshold is chosen for each level in the analysis.  Following Lega et
al.\,(1995) we calculate the wavelet transform for a random
distribution of particles where the number of particles and
``simulation volume'' are chosen to be the same as in the actual
simulation to be analyzed.  At each level, the threshold is chosen to
be five times the rms value of the wavelet coefficients.

At first glance, the $5\sigma$ threshold would seem to eliminate
virtually all chance associations of particles.  However, the
statistical fluctuations in the particle number are different in a
highly nonuniform halo simulation than in a random distribution.
Indeed, the wavelet transform maps not only substructure but also the
large scale structure of the main halo.  In any case, application of a
$5\sigma$ threshold is just the first step in producing a catalog of
substructures: Dynamical information is also used to remove unbound
(supposedly chance) associations of particles.  In the next section we
present results for different choices of the threshold which
demonstrate that the algorithm, taken as a whole, is relatively
robust.

The thresholding procedure leads to a binary representation of the
wavelet transform with a value $1$ ($0$) assigned to gridpoints where
the WAC is above (below) the threshold.  The next step it to identify
individual structures.  Initially, lists of structures (connected
region of above-threshold WACs) are generated for each level in the
transform.  These structures are identified by a method known as
segmentation analysis (Rosenfeld 1969; Lega et al.\,1995), the details
of which are presented in Appendix B.

Typically, the larger structures in a simulation lead to significant
WACs at multiple levels in the transform.  The spherically averaged
radial density profiles for halos and subhalos are well-approximated
by piecewise power-law functions of radius.  Therefore no single scale
can characterize a halo.  The upshot is that the combined list of
structures from all levels of a transform will contain numerous
redundancies.  For example, a single smooth subhalo (i.e., one devoid
of substructure) that is identified at the $n=4$ level will almost
certainly also be represented by entries in the $n=1,\,2,$ and $3$
levels.  These unwanted redundances are eliminated as follows: For a
given structure identified at the top level of the transform ($n=4$
for the case at hand), we remove a single entry, namely the one
closest to the center of the $n=4$ structure, from each of the lists
of structures at the lower levels.  The procedure is repeated for the
$n=3$ and $n=2$ lists.  The end result is an {\it irreducible} list of
subhalos and sub-subhalos where each distinct object is represented by
one and only one entry.

The final list of objects will contain not only gravitationally bound
systems but also unbound associations of particles (e.g., tidal
streams) and density enhancements that occur because of fluctuations
in the particle distribution.  Dynamical information enables us to
distinguish among these possibilities.  Our prescription for identifying
bound systems is described in Section 4.

\section{Substructure in a Cluster-Sized Halo}

Our substructure-finding algorithm is applied to the N-body
simulations of cluster-sized halos that appeared in Dubinski (1996) and
Ghigna et al.\,(1998).  These simulations were performed using
parallelized versions of the Barnes-Hut treecode (Barnes \& Hut 1986).
In each case, the main halo exhibits a high degree of substructure
with hundreds of subhalos.

In what follows, we describe the results from our analysis of the
Gigna et al.\,(1998) simulation (Figure 2).  A more extensive
discussion of these results as well as the results for the Dubinski
(1996) simulation can be found in Seymour (2000).  In the Gigna et
al.\,(1998) simulation, the virial radius of the main halo is $\sim
2\,{\rm Mpc}$.  There are approximately $5\times 10^5$ particles
inside this radius corresponding to a mass of $4.6\times
10^{14}\,M_\odot$.

We begin by constructing a $128^3$ pixelized map of the density field.
The DWT for levels $n=1-4$ are then calculated using the formulae in
Appendix A.  Two-dimensional projections of the wavelet coefficients,
constructed by selecting the maximum wavelet coefficient along each
`line-of-sight', are shown in Figure 3.  This figure illustrates the
manner in which substructure on different scales is captured in the
different levels of the DWT.  Note that the central object and many of
the larger satellites register strong signals in all four levels
illustrating the way in which the DWT captures the internal structure
of these systems.

The results of the segmentation analysis are shown in Figure 4 where
circles, superimposed on the projected particle distribution, are
drawn for the structures detected at each level.  The radii of the
circles corresponds to the size of the region found to be above
threshold (i.e., proportional to the cube root of the number of pixels
above threshold).  The fact that most of the large subhalos are
detected in multiple levels in the transform is reflected in the
appearance of concentric circles.  When this redundancy is removed
(see Section 2.3) 773, 223, 72, and 5 subhalos remain in levels $1-4$
respectively.

\section{Identifying Bound Systems}

The MRA described above identifies structures in the density field but
does not distinguish between particles that are members of a subhalo
and interlopers, i.e., particles that are passing through the subhalo
but are not associated with it in any true dynamical sense.  In
addition, some of the structures that we have identified are extremely
small and may in fact be chance associations that arise from Poisson
fluctuations in the density field.  Dynamical information makes it
possible to eliminate interlopers and also cull the list of structures
of unbound particle groups.

The standard working definition for a bound structure in an N-body
simulation is ``the largest set of particles that are {\it mutually}
gravitationally bound'' (see, for example, Bertschinger \& Gelb 1991;
Gigna et al 1998).  With this definition, the procedure for
identifying bound structures given the results of our wavelet and
segmentation analysis is as follows: (1) For each entry in our list of
structures, calculate its radius and center-of-mass.  For the radius,
we take $r=\left (3m/4\pi\right )^{1/3}\Delta a$ where $m$ is the
number of cells associated with the structure, i.e., connected cells
with WACs above the threshold.  Likewise, the center-of-mass is
calculated from the WACs that rise above the threshold.  (2) Next
calculate the total energy (kinetic plus potential) for each particle
within a distance $d<r$ from the structure's center of mass.  Note
that the kinetic energy is calculated in the rest frame of the
particles.  (3) The highest energy particle is identified and if its
energy is greater than zero, it is removed.  One then returns to steps
(2) and (3) with the proviso that in calculating the potential energy
and center of mass frame, only the remaining particles are used.
The procedure is repeated until the highest energy particle that
remains has negative energy.

In excluding unbound particles when calculating the gravitational
potential, one underestimates the (negative) potential energy of the
remaining particles and may therefore inadvertently remove bound
particles from a subhalo.  An alternative procedure is to recalculate
the center-of-mass velocity but {\it not} the potential energies each
time an unbound particle is removed.  In this way, all of the
particles in the neighborhood of the clump contribute to the
gravitational potential, as they should!

Figure 5 shows an example of a subhalo that is identified as a bound
system by the second method (interlopers included in the potential)
but rejected by the first method (i.e., no mutually gravitationally
bound (sub)system of particles was found).  The spatial distribution
of particles for the three Cartesian projections are shown in Figure
5a.  Note that only a small fraction of the particles (26 out of 532)
are identified as being members of the gravitationally bound system
thus explaining why, in this case, the background particles are so
important.  To check that we have indeed found a bona fide structure,
we show, in Figure 5b, the velocity space distribution of particles.
The bound clump forms a tight group in velocity space: Their velocity
dispersion is much smaller than that of all of the particles in this
region ($\sigma_{\rm bound}/\sigma_{\rm all} = 0.19$).  We can
estimate, from Poisson statistics, the probability that such a
velocity space clump will occur by chance.  The probability of finding
a single particle in the velocity-space region of the clump is $p =
\left (\sigma_{\rm bound}/\sigma_{\rm all}\right )^3\simeq 0.0063$ and
thus the probability of finding $n_b=26$ particles out of $n_t=532$ in
this region is $e^{-x}x^{n_b}/n_b!$ where $x = pn_t$.  Thus, the
probability of finding $n_b$ out of $n_t$ particles with velocity
dispersion $<\sigma_b$ in {\it any} region of velocity space is

\be{probability}
P = \frac{1}{p}\frac{e^{-x}x^{n_b}}{n_b!}~.
\ee

\noindent In the example of Figure 5, $P\simeq 3.7\times 10^{-13}$.

For many of the systems found by the second method, the case that they
are bona fide bound structures is not nearly so strong.  In
particular, for dense regions of the halo, it is not so unlikely to
find, by chance, small groups of particles that appear to
be bound.  Thus, we need some additional set of criteria to further
cull, from the list of structures, chance associations of particles.
The criteria we choose are as follows: For each bound system we
calculate the probability $P$ as above and also the fraction of local
particles $F=n_b/n_t$ that are found to be part of the bound system.
A bound clump is deemed to be genuine if either $P<10^{-4}$ or
$F>0.75$.  The first condition applies to bound systems moving through
dense regions of the halo (as in Figure 5) while the second condition
applies to relatively isolated systems in which $n_b\simeq n_t$ and
$\sigma_b\simeq \sigma_t$.  Our criteria are somewhat arbitrary and
other choices are possible.  For example, Gigna et al.\,(1998) only
consider groups of 16 or more particles.  This particle number cutoff
may be overly conservative as our algorithm identifies numerous
systems of fewer than 16 particles that are clearly clustered in both
velocity and configuration space.  Nevertheless, our results at the
low-mass end of the subhalo distribution may be suspect.  Clearly, the
relevant numerical experiment is to run a set of simulations with
idential initial conditions but with different numbers of particles.

The results for both methods of finding bound clumps are summarized in
Table 1.  With method 1, nearly three quarters of the level 1
structures and over one half of the level 2 structures are removed as
being ``unbound'' while with method two, less than half of the level 1
structures and less than one third of the level two structures are
removed.  The additional constraint eliminates (as potentially being
chance associations) an additional 100 level 1 structures.  

\bigskip

\begin{tabular}{ccccc} \hline
\multicolumn{5}{c}{$\sigma = 5$}\\ \hline
$n$ & $SA$ & $B1$ & $B2$ & $\overline{B2}$ \\ 
1    & 773     & 177     & 433     &  328  \\
2    & 223     & 119     & 158     &  158  \\ 
3    & 72     &   72   & 72     &     72   \\ 
4    &  5    &    5  & 5     &        5    \\ \hline
\end{tabular}

\bigskip

We have repeated the procedure outlined above using both $4\sigma$ and
$6\sigma$ thresholds and the results are shown in Tables 2 and 3.  As
expected, the number of structures identified by our wavelet and
segmentation analysis using a $4\sigma$ threshold is higher than the
number identified with a $5\sigma$ threshold and likewise the number
identified using a $6\sigma$ threshold is lower.  However, the
unbinding procedure seems to identify the greatest number of
structures when a $5\sigma$ threshold is used.  The unbinding
procedure begins with a region specified by the segmentation analysis
where it can search for a bound object.  If the region is much larger
than the clump, and the clump center-of-mass velocity is much
different from the center-of-mass velocity of the background, the
unbinding procedure may miss the clump altogether.  These results
illustrate a truism common to all substructure-finding algorithms: On
an object by object basis near the mass resolution of the simulation
subhalo identification is often ambiguous.  That is, one is likely to
miss some bound clumps while including, in the catalog of structures,
some chance associations.  However, as we demonstrate below, the
statistical properties of the subhalo population (namely fraction of
main halo in bound clumps) appears to be insensitive to the choice of
threshold.

\bigskip

\begin{tabular}{ccccc} \hline
\multicolumn{4}{c}{$\sigma = 4$}\\ \hline
$n$ & $SA$ &  $B2$ & $\overline{B2}$ \\ 
1    & 820     & 335    &  244  \\
2    & 212     & 200     &  187  \\ 
3    & 77     & 77     &     77   \\ 
4    &  10    &    10    &        10    \\ \hline
\end{tabular}

\bigskip

\begin{tabular}{ccccc} \hline
\multicolumn{4}{c}{$\sigma = 6$}\\ \hline
$n$ & $SA$ &  $B2$ & $\overline{B2}$ \\ 
1    & 439     & 313    &  240  \\
2    & 134     & 134     &  134  \\ 
3    & 63     & 63     &     63   \\ 
4    &  3    &    3    &        3    \\ \hline
\end{tabular}

\bigskip

We conclude this section by considering a region of the simulation
$800\, {\rm kpc}$ on a side ($\sim 1.5\%$ of the simulation volume)
centered on a large subhalo that has been detected at level 4 of the
DWT.  The upper left panel of Figure 6a displays all of the particles
in this region while the upper right panel displays only those
particles gravitationally bound to the main subhalo.  We have also
identified twelve smaller subhalos within the region defined by the
main subhalo.  These are shown in the lower left and lower right
panels (levels 2 and 1 of the analysis respectively).  (Many of the
clumps that appear in the upper left panel are in the foreground or
background of the main subhalo and therefore are not shown.)  One of
these small subhalos, indicated by the arrow, is gravitationally bound
to the main subhalo.  In addition, a second subhalo appears to be
associated dynamically with the main subhalo.  The remaining ten
subhalos are interlopers as is evident from Figure 6b where the 12
small subhalos are plotted in velocity space.  The circle represents
the velocity dispersion of the large subhalo.  Note that the relative
velocities of the subclumps are typically much higher than their
internal velocities.

The gravitationally bound subhalo is an example of a third level of
substructure in that its constituents can be regarded as members of
three distinct systems.  Examples of this type are rare, a result that
is not surprising given that the systems which constitute the second
level of substructure (the large subhalos orbiting the main halo) are
consist of only a few thousand particles and therefore subject to the
purely numerical overmerging problem.

\section{Substructure Statistics}

Our subhalo detection algorithm has identified over 500 subhalos and
it is therefore possible to study, in a statistical sense, their
characteristics.  In Figure 7, for example, the rms internal velocity
dispersion of the clumps are plotted as a function of their position
within the main halo.  As noted in Ghigna et al.\,1998, a wide range
of clump velocities are found throughout the main halo though there
does appear to be a trend toward larger internal velocities for
subclumps closer to the center of the main halo.  This result may be
an indication that subhalos are heated by the tidal field of the main
halo.  In addition, there appears to be a slight enhancement of
subclumps with large rms velocities at $r\simeq 1.5\,{\rm Mpc}$.
These subhalos are probably associated with the large subhalo evident
in Figure 2 at $x\simeq 1.2\,{\rm Mpc},~y\simeq 0.1\,{\rm Mpc}$.  In
Figure 8, the masses of the subhalos (shown in terms of particle
number) are plotted as a function of rms velocity.  One finds a
reasonably tight correlation between $M$ and $v_{\rm rms}$ with
$M\propto v_{\rm rms}^{\alpha}$ where $\alpha\simeq 3$.  The
correlation appears to hold over nearly 3 orders of magnitude in mass.
This result is to be compared with Ghigna et al.\,1998 (see their
Figure 20) who find $\alpha\simeq 3-3.4$ over approximately two orders
of magnitude in mass.

In Figure 8, there appears to be a distinct population of objects that
have, for fixed mass, velocities a factor of 3-5 greater than those in
the main distribution.  These objects are found preferentially in the
inner regions of the main halo once again suggesting that they are
heated by the tidal fields of the main halo.  This result brings us to
Figure 9 where we plot the virial ratio, $2T/|W|$ as a function of
position in the main halo.  $T$ and $W$ are respectively the total
kinetic and potential energies of the bound system.  $T$ is simply the
sum of the kinetic energies of the particles in the system.
Calculation of $W$ is somewhat more involved since a sum of the
potential energies of the individual bound particles double counts for
pairs of bound particles.  The formula for $W$ is:

\be{potential}
W = -\sum_{i\in \{B\}}\left (
\frac{1}{2}\sum_{j\in \{B\};\,j\ne i} 
\frac{m_i m_j}{|{\bf r_i-r_j}|} + 
\sum_{j\in \{U\}}\frac{m_i m_j}{|{\bf r_i-r_j}|}\right )
\ee

\noindent where $\{B\}$ and $\{U\}$ refer respectively to the sets of
bound and unbound particles or interlopers.  The population of
subhalos exhibit a large scatter in the virial ratio about the
equilibrium value of $1$.  This result implies that the halos are, for
the most part, not fully virialized, perhaps because that are
constantly being disturbed (if not disrupted) by the tidal field of
the main halo.  Indeed, there appears to be a trend toward a larger
dispersion in the virial ratio toward the center of the main halo.

Finally, in Figure 10, we plot the fraction of the virial mass
contained in subhalos of a given size.  We find that in total, $12\%$
of the halo is bound in subhalos in good agreement with results from
Gigna et al.\,(1998).  The figure includes results for different
choices of the threshold.  The fact that the results do not change much
is an indication of the robust nature of the algorithm.

\section{Conclusions} 

Recent advances in numerical cosmology lead to the conclusion
that the more particles one uses in an N-body simulation, the more
substructure one finds in dark matter halos.  This result is
consistent with the picture that halos form through hierarchical
clustering since an increase in the number of simulation particles
translates into an increase in the dynamic range of the hierarchy
accessible to the simulation.  What is perhaps surprising is the
extent to which substructure survives.

The identification of substructure inside dark matter halos is a
nontrivial task.  While large subhalos are easily located by eye,
small subsystems, and especially ones inside dense regions of the
halo, can be difficult to spot.  Moreover, there can be many levels
of substructure.

The substructure-finding algorithm presented in this paper is
intrinsically hierarchical and therefore perfectly suited to the task
at hand.  The DWT analyzes the density field at different resolutions
while segmentation analysis produces a catalog of structures at each
level of the transform.  Further analysis is required to eliminate
redundances and cull the catalog of unbound systems.

One advantage of a wavelet-based analysis is that it is insensitive to
a constant or linearly varying background density.  This feature makes
it possible to pick out structures such as the one shown in Figure 5.
Note that the density enhancement for this structure is only $\sim
5\%$.  To pick such a structure out using FOF would require an extremely
finely tuned linking parameter.

Our results indicate that the MRA algorithm is competitive with FOF
and DENMAX/SKID, the two most widely used methods.  In particular, our
results for the number of subhalos found, the fraction of mass in
subhalos, and subhalo statistics are consistent with those of Gigna et
al.\,(1998).  A detailed comparison of the three methods for a
sequence of simulations performed at increasing particle resolution is
required to determine if there is any true advantage of one method
over the others.

\acknowledgements{We are grateful to B. Moore and J. Dubinski for
providing us with their simulations.  We thank A. Babul, J. Navarro,
R. Henriksen, K. Perrett, J. Stadel, and D. Stiff for useful
conversations.  LMW would like to thank the University of Chicago and
the Canadian Institute for Theoretical Astrophysics for their
hospitality during recent visits.  This work was supported, in part,
by the Natural Science and Engineering Research Council of Canada.}

\clearpage

\appendix

\section{WACs for the \bfAtrous}
 
In this appendix, we provide the details required to implement the
\iatrousp.  Consider a one-dimensional function
$f(x)$.  The highest resolution sample of $f$ is constructed 
from Eq.\,\EC{smooth1} or equivalently

\be{step1}
f_{0,i} = \frac{1}{6}\left (f\left (x_{i-1}\right )+
4f\left (x_{i}\right )+f\left (x_{i+1}\right )\right )~.
\ee

\noindent $f_1$ can be calculated from $f_0$ using 
Eqs.\,\EC{smooth2} and \EC{recursion2} as follows:

\bea{step2}
f_{1,i} &=& \frac{1}{2}\sum_j \phi\left (\frac{x_i-x_j}{2}\right )
\,f\left (x_j\right ) \\
&=& \frac{1}{16}\sum_j \sum_k\,C^4_{2-k}\,
\phi\left (x_i-x_j-x_k\right )f\left (x_j\right )\\
&=& \frac{1}{16}\sum_k\,C^4_{2-k}\,f_{0,i-k}~.
\eea

\noindent Likewise, the coefficients corresponding to smoother
samples of $f$ can be calculated recursively:

\be{step3}
f_{n,i} = \frac{1}{16}\sum_k \,C_{2-k}^4\,f_{n-1,m}
\ee

\noindent where $m\equiv i-k2^{n-1}$.
The WACs are calculated by a simple subtraction
(Eq.\,\EC{details}.

Next, we consider a three-dimensional density field $\rho\left
(x,y,z\right )$.  As in the one-dimensional case, the first step is to
compute the auxiliary coefficients, here written as $c(n,i,j,k)$.  The
$n=0$ coefficients are computed as follows:

\be{step13D}
c(1,i,j,k) = \sum_{i',j',k'=-2}^2\rho(i',j',k')
\phi\left (i'-i\right )
\phi\left (j'-j\right )
\phi\left (k'-k\right )
\ee

\noindent where we abbreviate $x_i=i\Delta a$ as $i$.  The $n>1$
coefficients are given by

\be{step23D}
c(n,i,j,k) = \sum_{i',j',k'=-2}^2
h\left (i'\right )h\left (j'\right )h\left (k'\right )
c\left (n-1,i-i'2^{n-1},j-j'2^{n-1},k-k'2^{n-1}\right )
\ee

\noindent where $h(i)\equiv \frac{1}{16}C^4_{2-i}$.  Finally, the
wavelet coefficients are found by performing the subtraction

\be{step33D}
w(n,i,j,k) = c(n,i,j,k)-c(n+1,i,j,k)
\ee

\section{Segmentation Analysis}

This appendix provides an outline of the segmentation analysis routine
used in the paper.  Further details can be found in Rosenfeld (1969)
and Lega et al.\,(1995).  The algorithm is introduced by way of a
worked example in two spatial dimensions.  The extension to three
dimensions is straightforward.

Figure 11a presents the results for a single level of a mock wavelet
transform of a two-dimensional density field.  The number in each cell
represents the WAC (for convenience, rounded off to the nearest whole
number).  Application of a threshold between 4 and 5 leads to Figure 11b.

The core of the segmentation algorithm is the {\it scan}.  The scan
identifies connected segments of non-zero elements along each of the
two (or three, for the actual data) Cartesian directions.  Starting in
upper left corner of the data array, the first scan proceeds along
rows successively from top to bottom labelling each segment 
of $1's$ by a unique
integer.  The result of the left-to-right scan is shown in 
Figure 11c.

Next, one scans from top to bottom along successive columns.  In this
scan one assigns the lowest integer contained in each segment.  The
result is shown in Figure 11d.  For a three-dimensional data set,
there is an additional to be performed.

Iterating the above procedure (i.e., alternating between horizontal
and vertical scans) will eventually yield the desired result: a unique
label for each connected structure.  However, for large data sets and
odd-shaped structures, the computation time can become prohibitively
long.  For a data set containing $N^2$ elements, each scan takes
$O(N^2)$ operations ($N$ operations along each row times $N$ columns
or vice versa).  One can imagine particularly perverse examples (for
example, staircase-like structures) in which one requires $O(N^2)$
scans, in other words, $O(N^4)$ computations.  In three dimensions,
one could in principle require $O(N^9)$ computations which is
unreasonable for even modest values of $N$.

One can obviate the need for repeated scans by means of the following
scheme.  The scan is performed once in each of the Cartesian
directions.  By virture of the scanning process, each cluster of
connected segments is described by a unique set of integers.  That is,
no cluster of connected segments contains values of another cluster.
In our example, the cluster on the left contains values $1,\,4,\,6,\,$
and $8$ while the cluster on the right contains values $2,\,7,\,$ and
$11$.  Our goal is to group together the integers represented in each
cluster.  This step is accomplished by first constructing a dictionary
as follows: For each non-zero element in the array, check all non-zero
neighboring elements and list the element and its neighbor if they are
different.  Note that the last direction scanned (in our example, the
vertical direction) need not be check since all non-zero neighbors are
necessarily the same.  The dictionary is then pruned for
duplicates (Figure 12a).

The remaining step is to sort the dictionary into connected pairs.
First, one picks out pairs that have the same number in either the
first or second position.  The sort for the cluster on the left of our
data is shown in Figure 12b.  We then relabel the second digit of
those pairs selected by the sort (Figure 12c).  The group is then
pruned to eliminate redundancy (Figure 12d) to yield a table that
identifies all of the integers for cluster 1 with the single integer
4.  Figure 12e shows the result when the remaining dictionary is
sorted.

{}

\clearpage

\figcaption[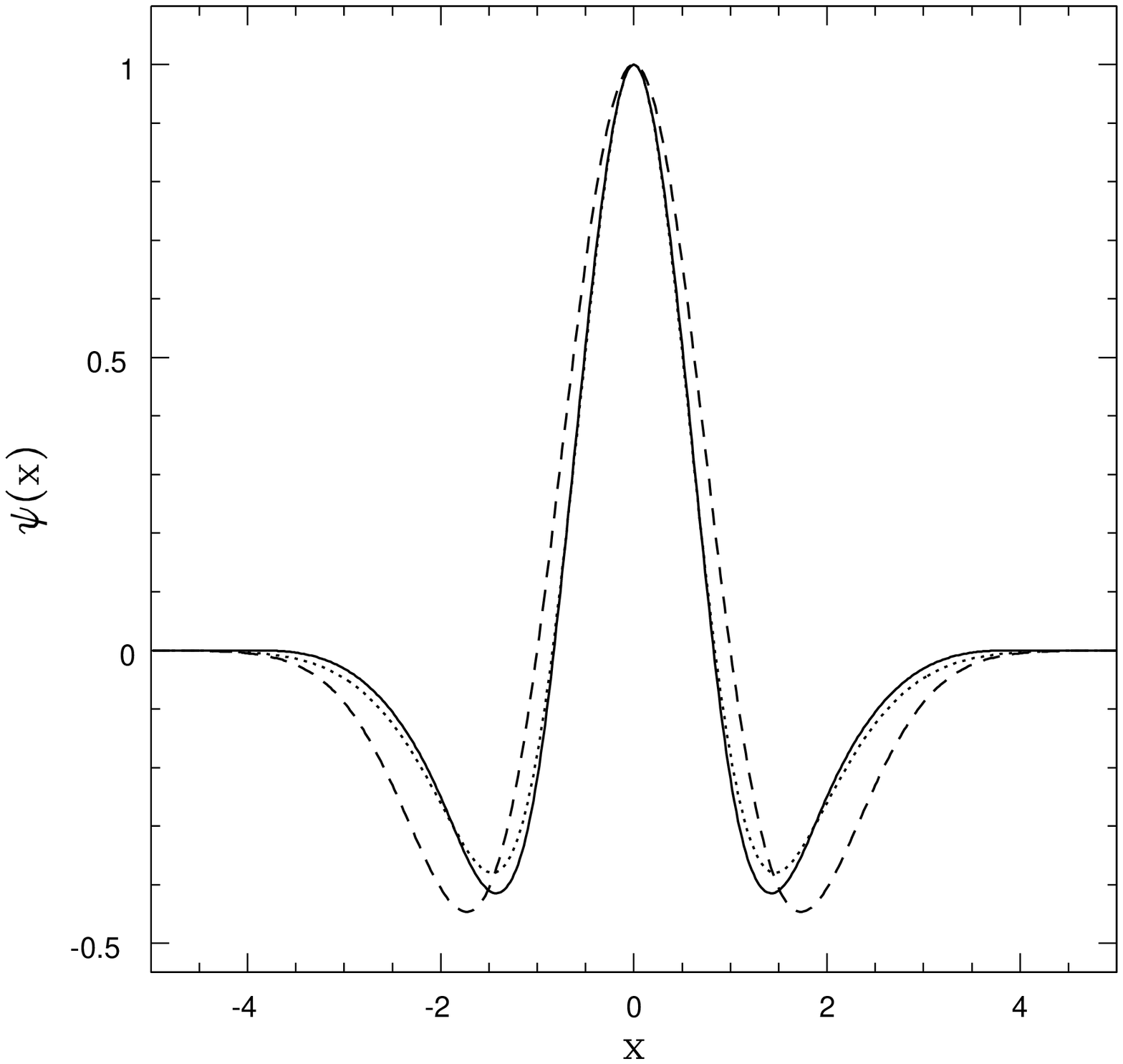]{Mother wavelet for the \iatrous in
three dimensions along one of the principle axes (solid line) and
along the diagonal $x=y=z$ (dashed line).  The wavelet has been
normalized to have a peak value of $1$.  For comparison, the Mexican hat
wavelet is also shown (dotted line).
\label{f1}}

\figcaption[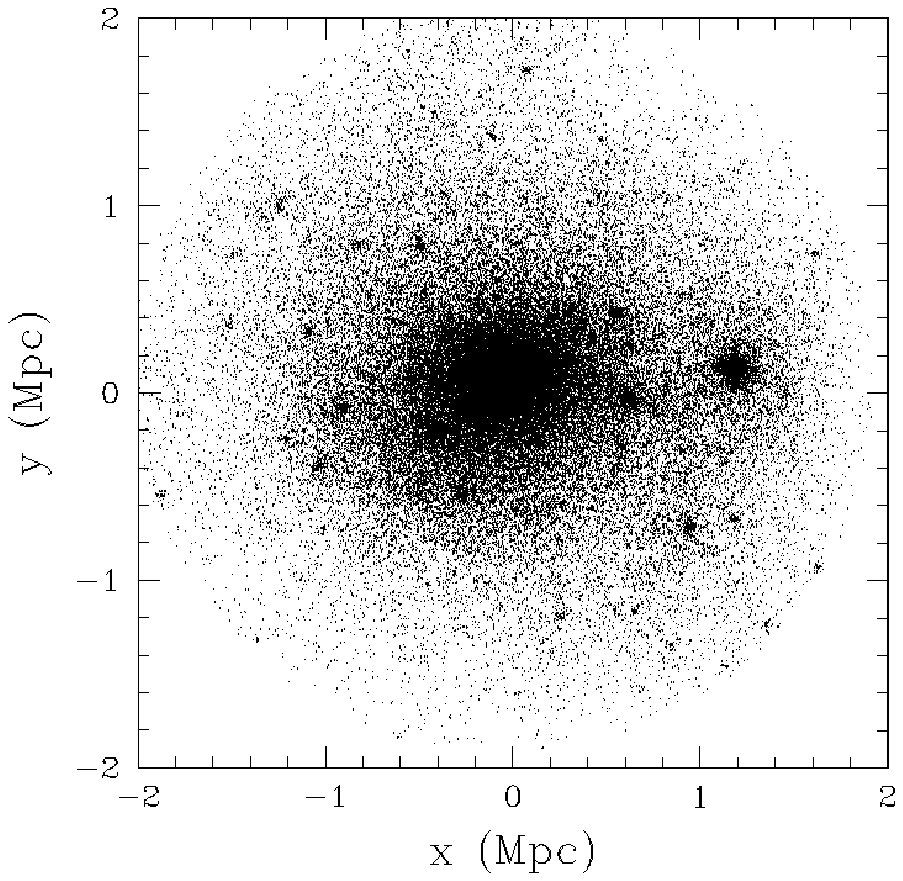]{Particle distribution for the halo analyzed
in this paper.  Only particles within $2\,{\rm Mpc}$, the virial 
radius.  For clarity, only one in five particles are plotted.  The data
is from Gigna et al.\,(1999).
\label{f2}}

\figcaption[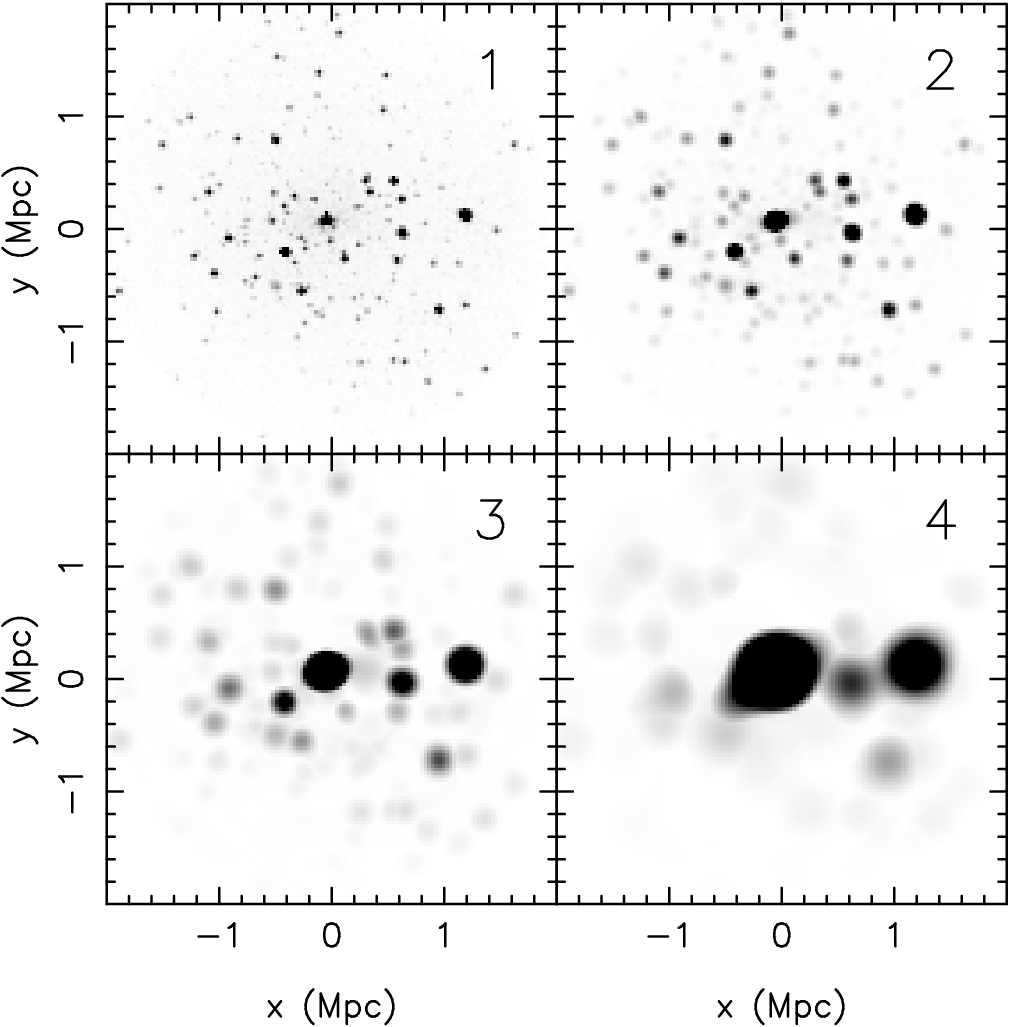]{Grayscale maps of the wavelet coefficients for
levels 1-4 of the wavelet transform.  The level is indicated in the 
upper right corner of each panel.  Shading corresponds to the largest
wavelet coefficient at a given level along each line of sight.
\label{f3}}

\figcaption[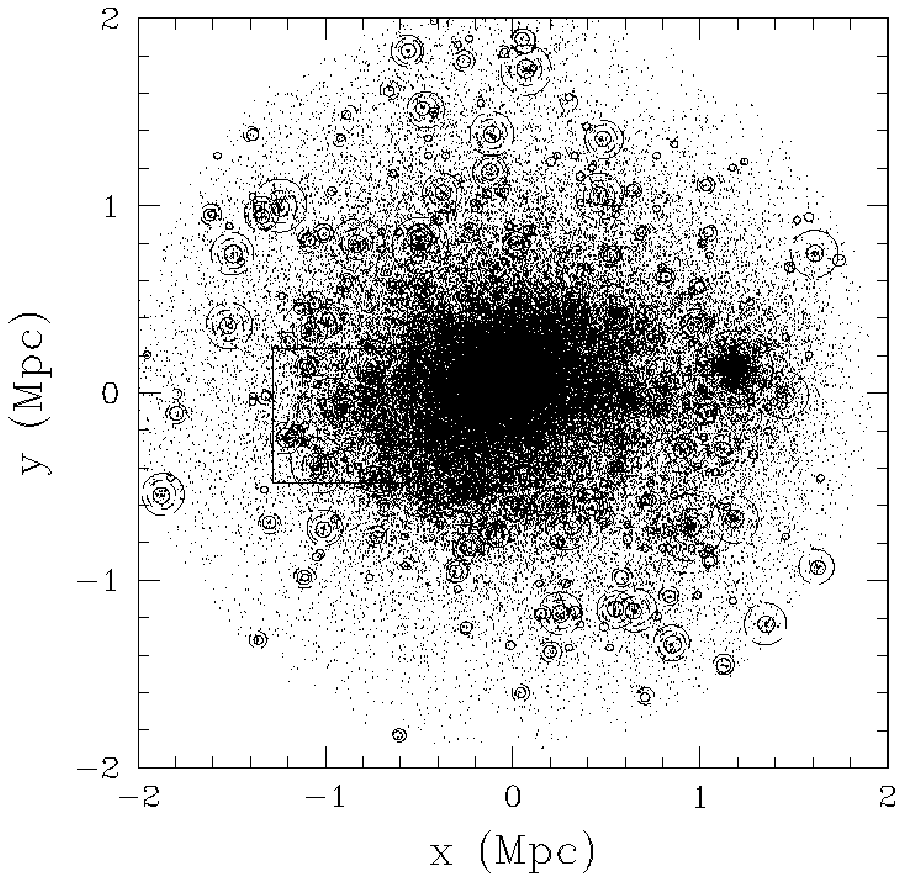]{Same as Figure 2 but with circles superimposed 
on the particle distribution corresponding to regions of pixels above
threshold.  The different linetypes correspond to the different levels
in the wavelet transform: Level 1 -- solid line; Level 2 -- dashed
line; Level 3 -- dot-dashed line; Level 4 -- dotted line.  The box to
the left of center in the figure refers to Figure 6.
\label{f4}}

\figcaption[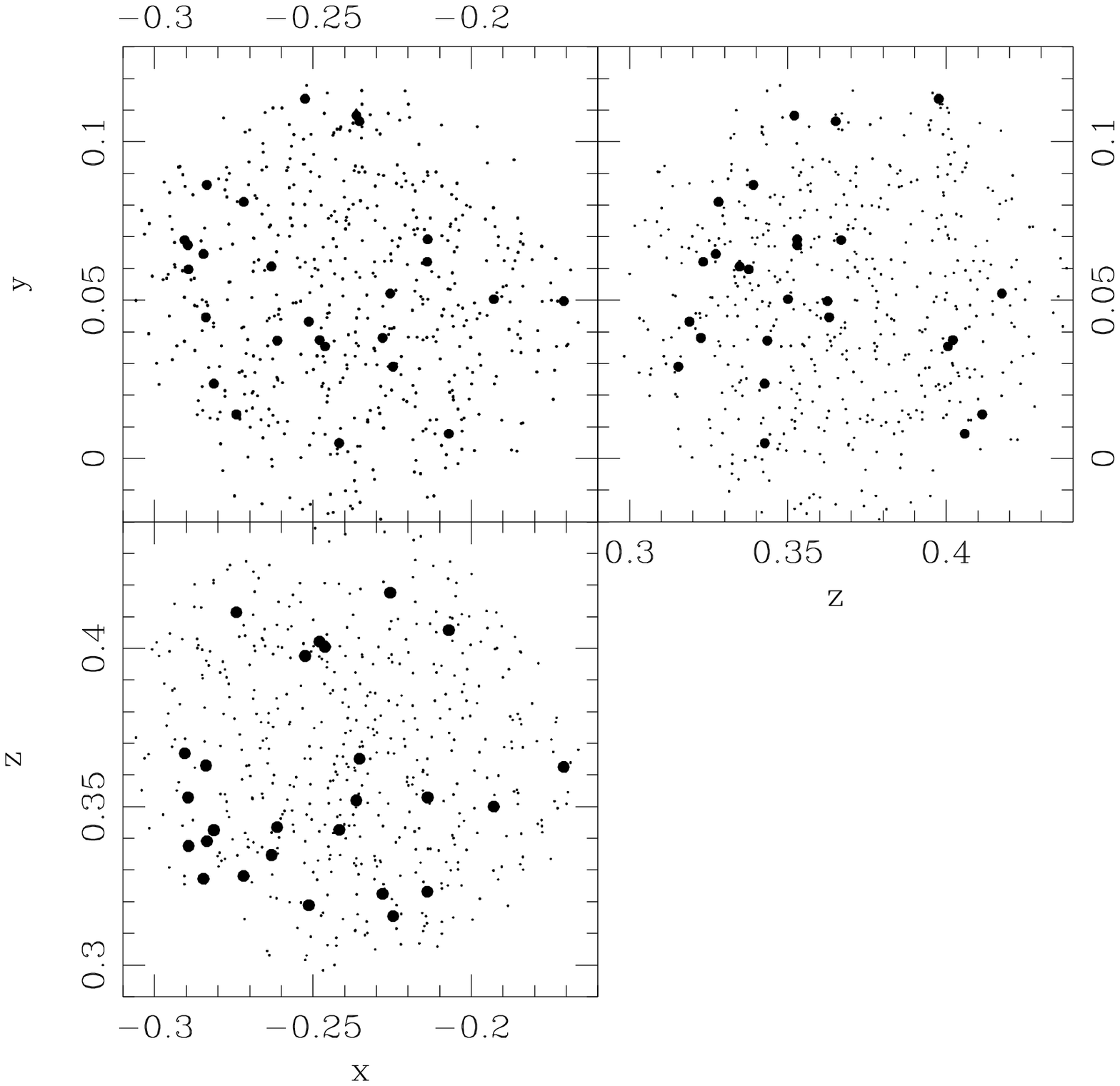,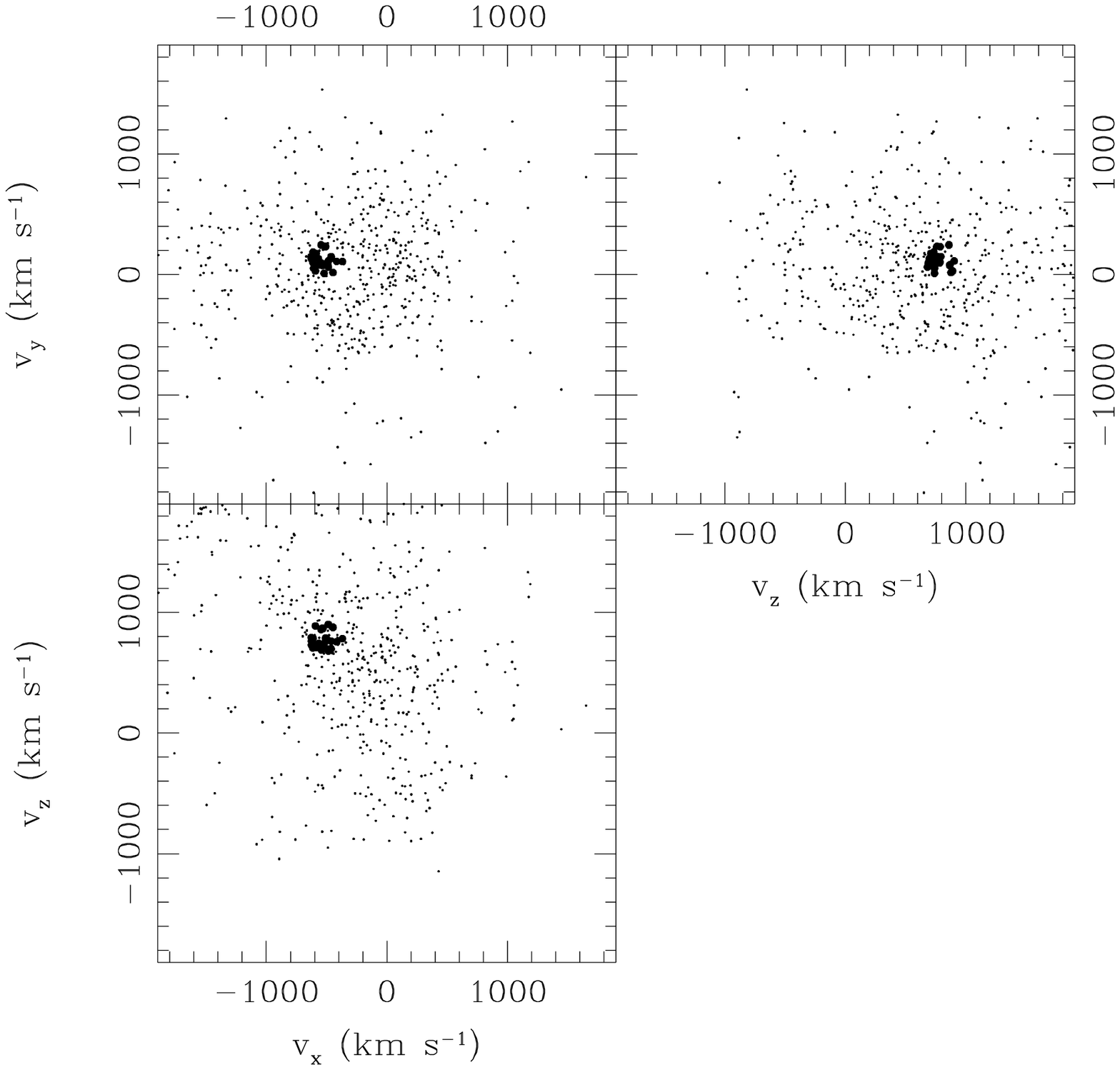]{Projections in position and
velocity space of the gravitationally bound clump discussed in the text.
Only particles in the region of the clump are shown.  Small dots are
interlopers.  Large dots are gravitationally bound particles.  Figure 5a
shows the distribution of particles in position space while Figure 5b
shows their distribution in velocity space.}

\figcaption[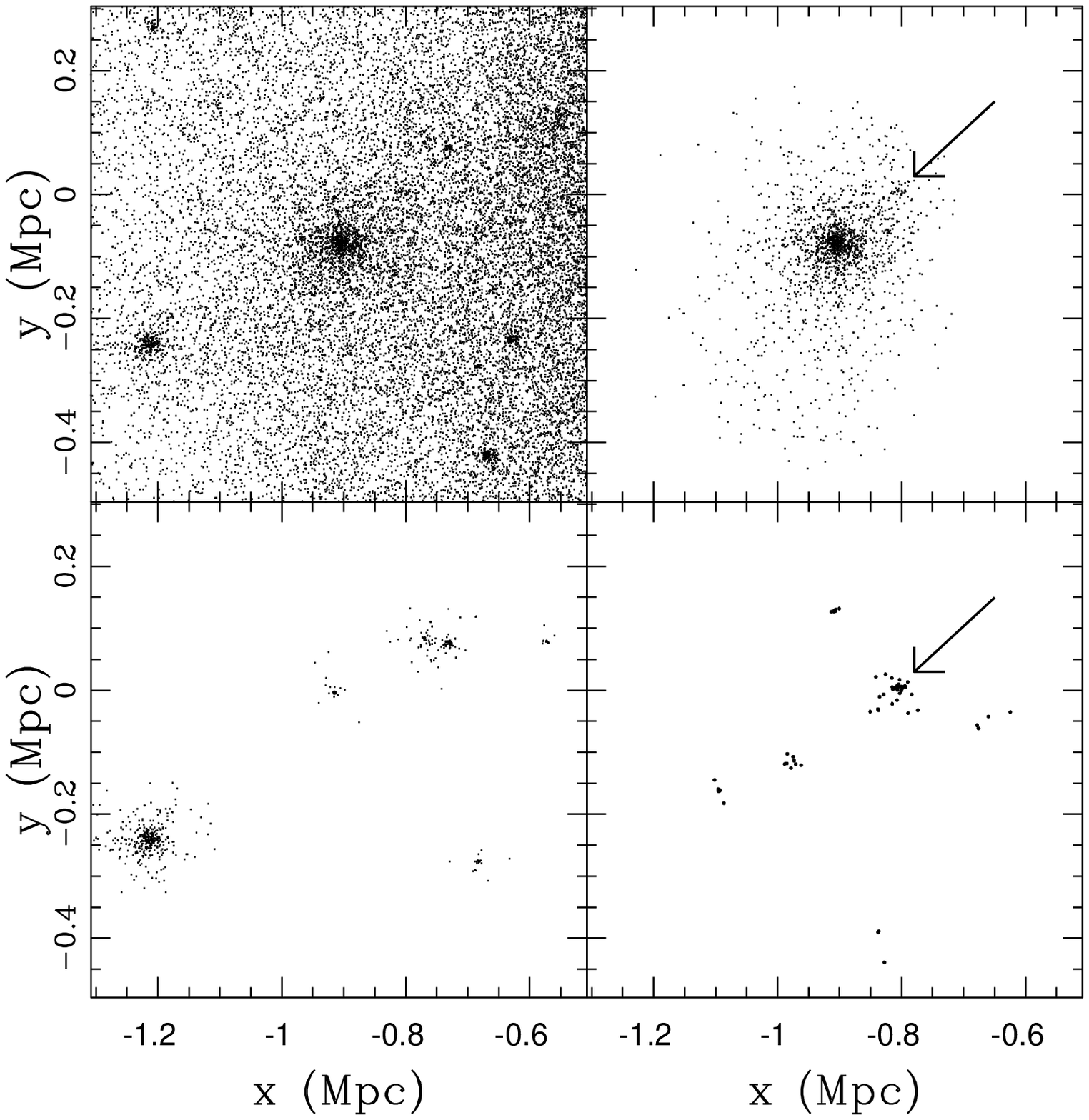,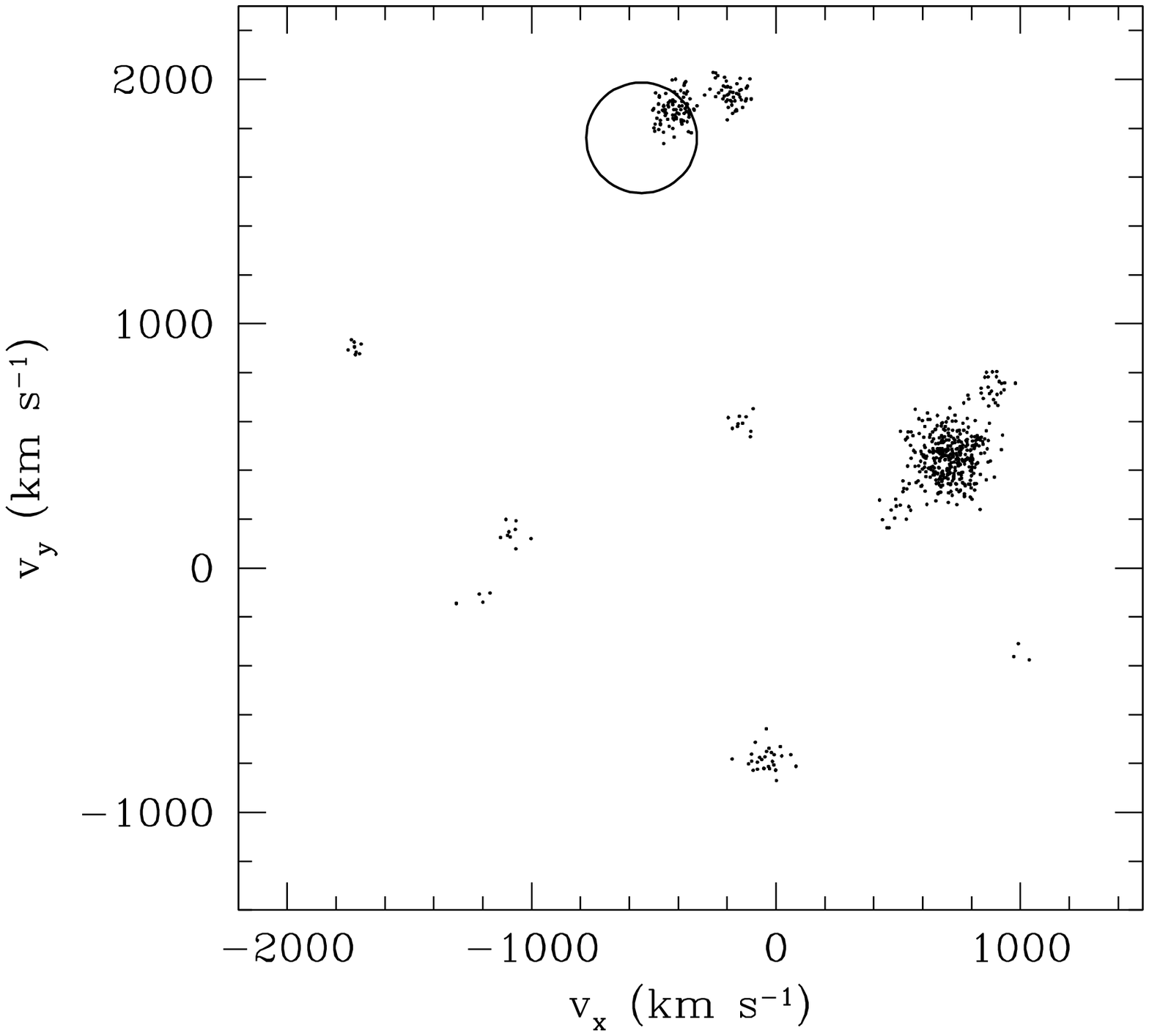] {Close-up view of a region
$200\,{\rm kpc}$ in size centered on a large subhalo.  Figure 6a:
Upper left panel shows all particles in this region.  Upper right
panel shows only those particles gravitationally bound to the main
subhalo.  Lower left and lower right panels (levels 2 and 1
respectively) show subclumps whose spatial extent overlaps with the
large subhalo.  The arrows in the upper and lower righthand panels
points to a subhalos that is gravitationally bound to the large
subhalo.  Figure 6b: Velocity space view of the 12 level 2 and level 1
subhalos shown in Figure 5.  The circle represents the internal
velocity dispersion of the main subhalo.
\label{f6}}

\figcaption[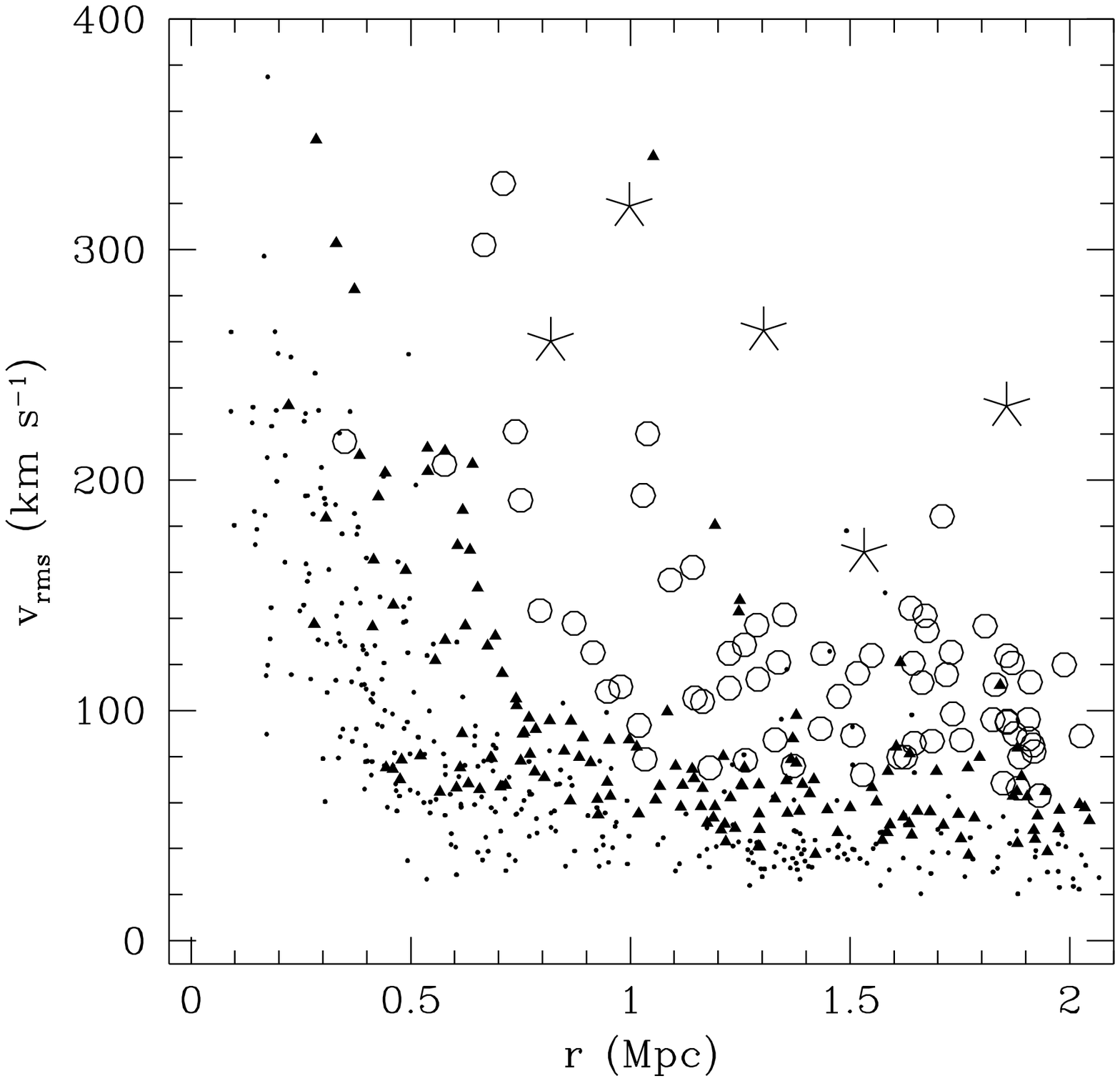]{RMS velocity versus radial position for the 
subhalos found in this analysis.  The type of symbol indicates the
level at which a particular subhalo is found:  Level 1 -- dot; Level 2
-- triangle; Level 3 -- open circle; Level 4 -- star.
\label{f7}}

\figcaption[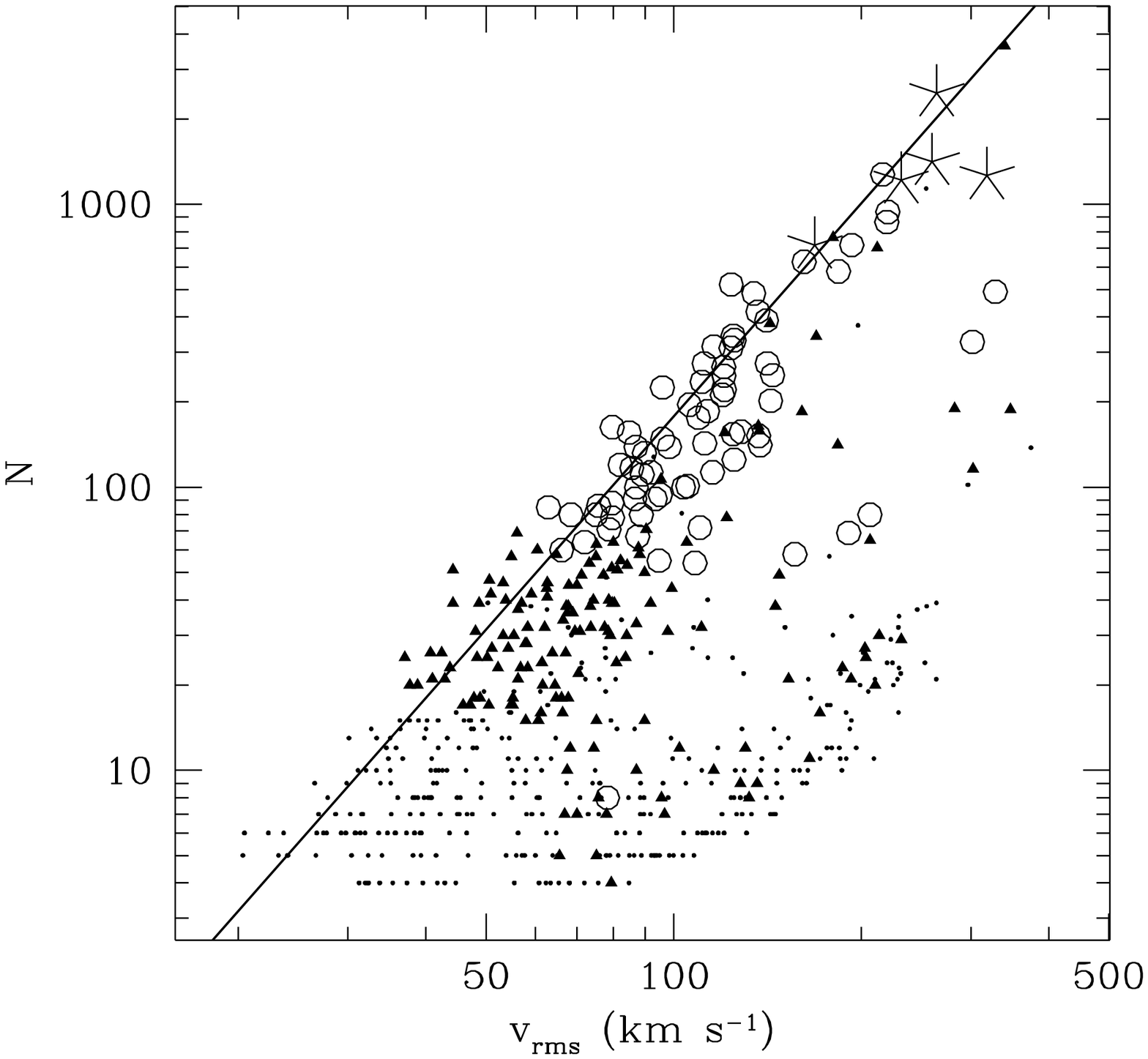]{Number of particles versus RMS velocity for
the subhalos found in this analysis.  Each particle has a mass
$m_p=8.56\times 10^8\,M_\odot$.  Symbols are the same as in Figure 7.
\label{f8}}

\figcaption[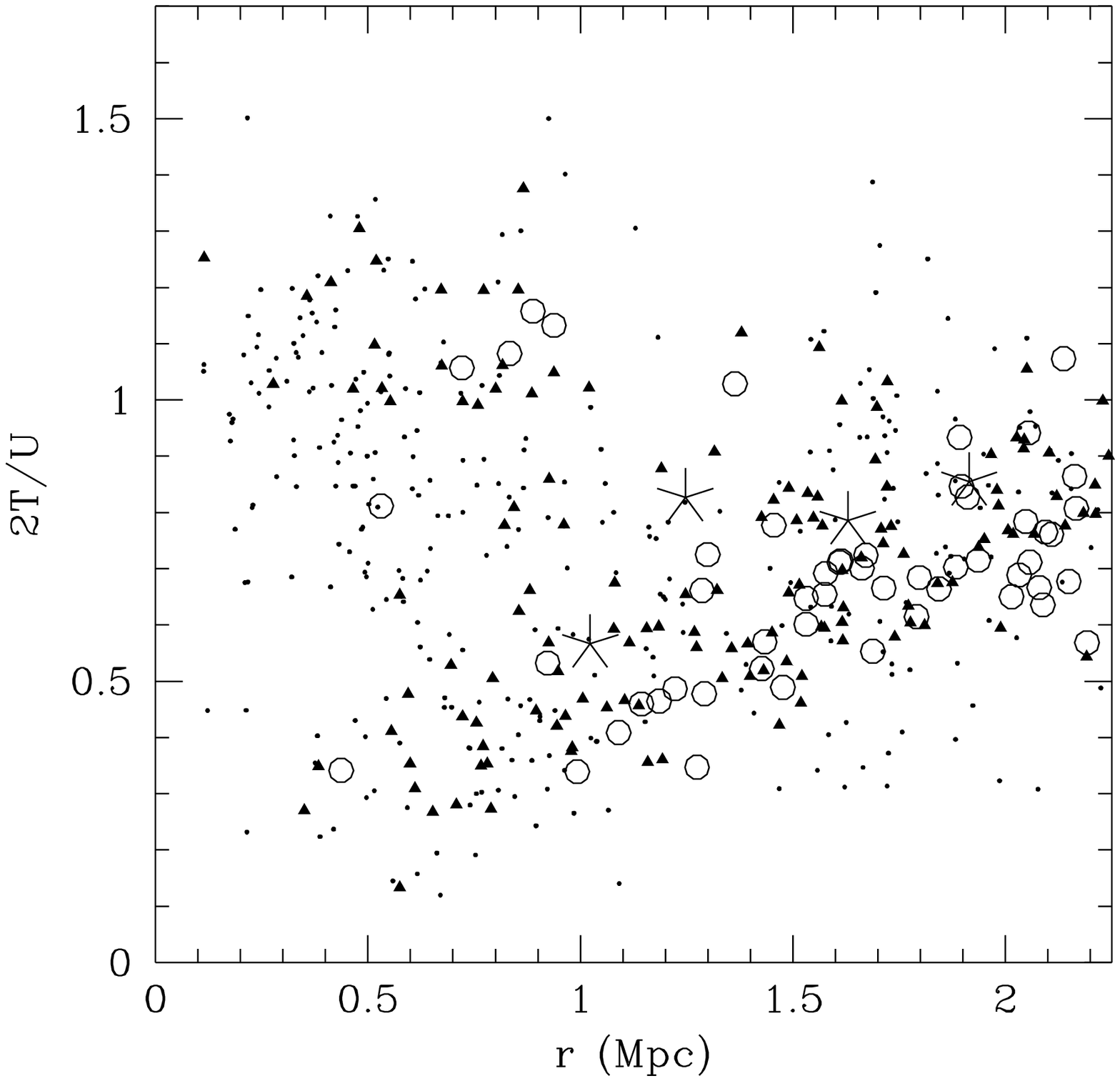]{Virial ratio $2T/U$ as a function of radius.
Symbols are the same as in Figure 7.
\label{f9}}

\figcaption[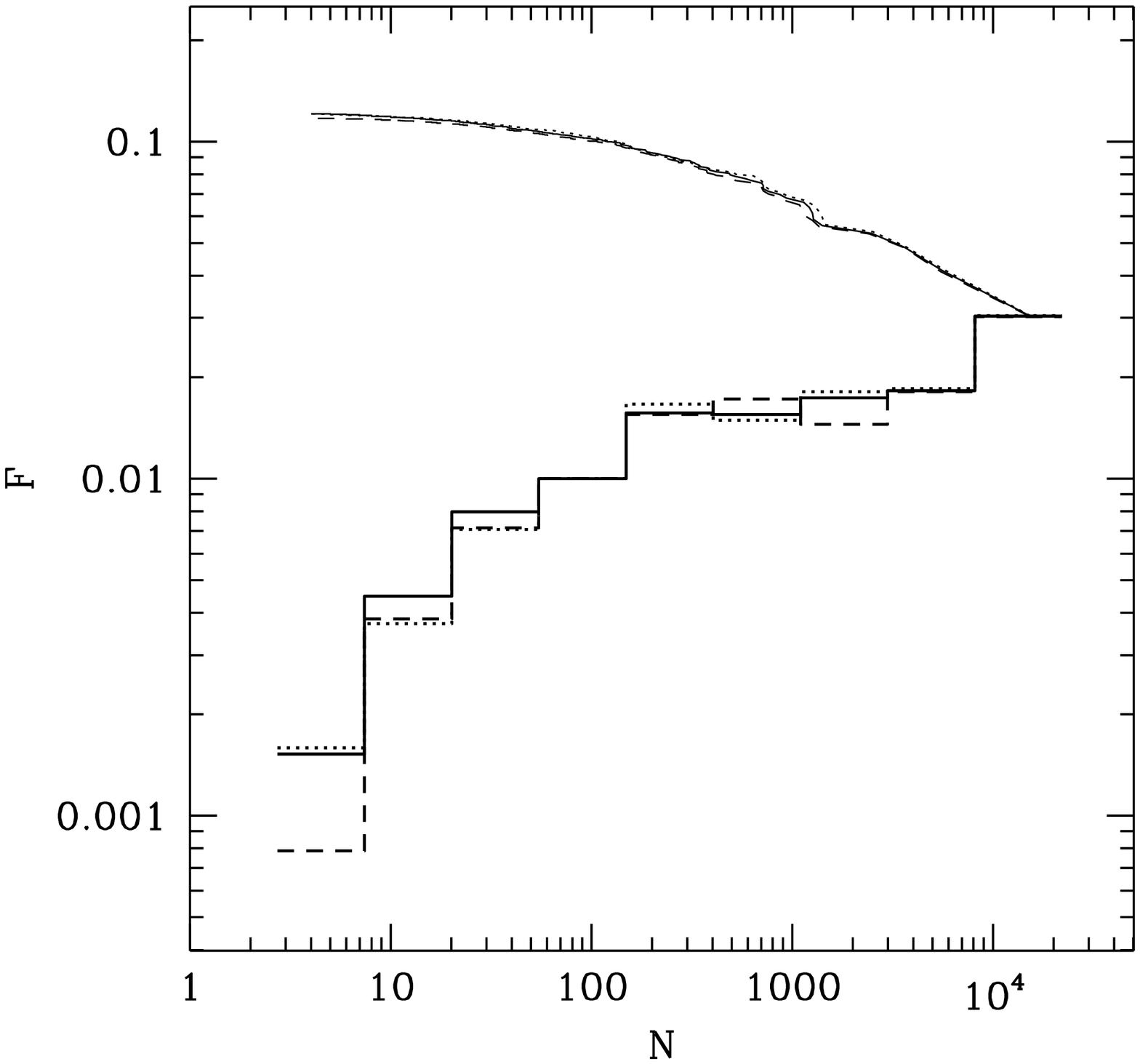]{Fraction of particles within the virial
radius of the main halo that are a member of a subhalo.  Heavy lines
give a histogram of the halo fraction as a function of subhalo mass.
Mass is shown in terms of particle number ($1$ particle $=\,9.1\times
10^{8}\,M_\odot$).  Light lines give the cumulative fraction for
particles within subhalos above a certain mass.  Solid line gives the
results for a $5\sigma$ threshold; Dotted line gives the results for a
$6\sigma$ threshold; Dashed line gives results for a $4\sigma$
threshold.}

\figcaption[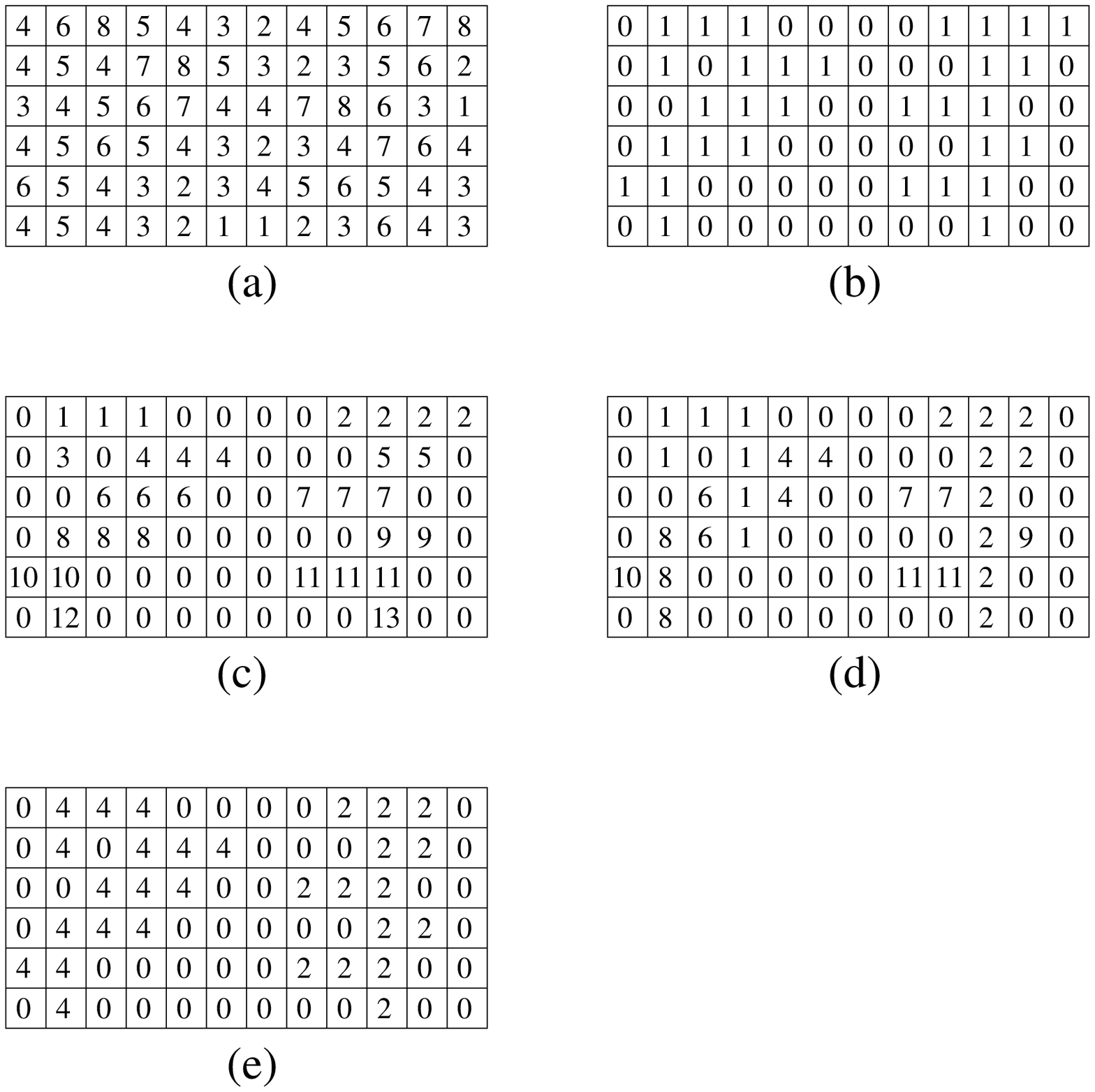]{Illustration of the segmentation analysis
used in the text.  Figure 11a: Mock two-dimensional data set.  Numbers
represent WACs rounded to the nearest integer.  Figure 11b: Data set
after a threshold of 4.5 has been applied.  Figure 11c: Result of
scans along rows from left to right.  Figure 11d: Result after a scan
along columns from top to bottom.  Figure 11e: Result after the
dictionary has been applied (see Figure 12).}

\figcaption[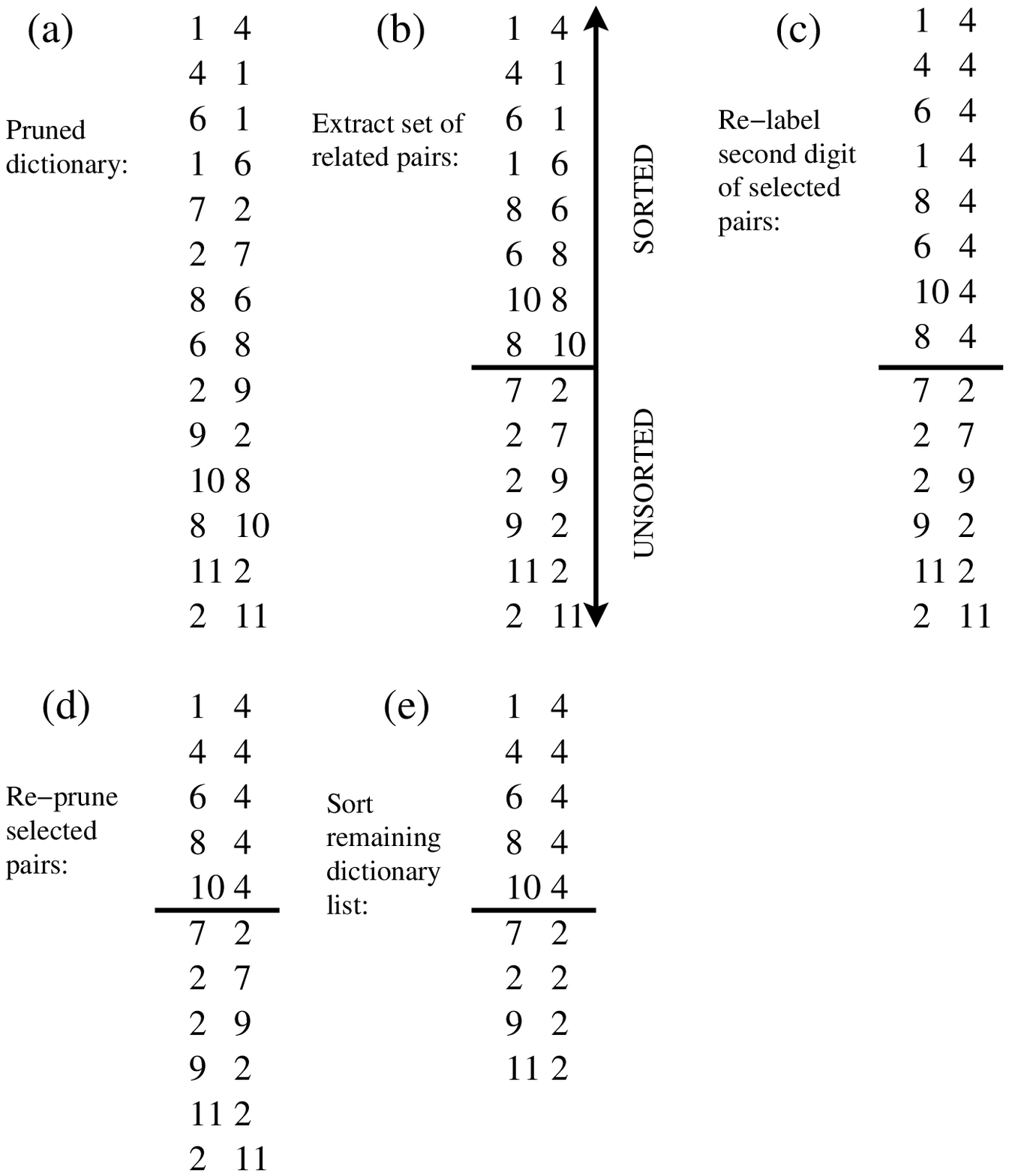]{Algorithm to yield unique labels for each of
the structures in the data set.  Figure 12a: Pruned dictionary of neighboring
pairs of adjoining differing pairs of (non-zero) elements.  Figure 12b: Result
after the dictionary has been sorted to isolate pairs belonging to the 
structure on the left hand side of the data set.  Figure 12c: Second digit of
each pair is relabelled to the second digit of the first pair.  Figure 12d:
Re-prune list for left-hand structure.  Figure 12e: Process is repeated for
structure on the right-hand side of the data set.}

\clearpage

\begin{figure}
\plotone{figure1.ps}
\label{figure1}
\end{figure}

\begin{figure}
\begin{center}
\plotone{figure2.ps}
\label{figure2}
\end{center}
\end{figure}

\begin{figure}
\begin{center}
\plotone{figure3.eps}
\label{figure3}
\end{center}
\end{figure}

\begin{figure}
\begin{center}
\plotone{figure4.ps}
\label{figure4}
\end{center}
\end{figure}

\begin{figure}
\begin{center}
\plotone{figure5a.ps}
\label{figure5a}
\end{center}
\end{figure}

\begin{figure}
\begin{center}
\plotone{figure5b.ps}
\label{figure5b}
\end{center}
\end{figure}

\begin{figure}
\begin{center}
\plotone{figure6a.ps}
\label{figure6a}
\end{center}
\end{figure}

\begin{figure}
\begin{center}
\plotone{figure6b.ps}
\label{figure6b}
\end{center}
\end{figure}

\begin{figure}
\begin{center}
\plotone{figure7.ps}
\label{figure7}
\end{center}
\end{figure}

\begin{figure}
\begin{center}
\plotone{figure8.ps}
\label{figure8}
\end{center}
\end{figure}

\begin{figure}
\begin{center}
\plotone{figure9.ps}
\label{figure9}
\end{center}
\end{figure}

\begin{figure}
\begin{center}
\plotone{figure10.ps}
\label{figure10}
\end{center}
\end{figure}

\begin{figure}
\begin{center}
\plotone{figure11.ps}
\label{figure11}
\end{center}
\end{figure}

\begin{figure}
\begin{center}
\plotone{figure12.eps}
\label{figure12}
\end{center}
\end{figure}

\end{document}